\documentclass{aa}
\input epsf
\begin{document}
   \thesaurus{06     
              (08.03.4;  
               08.05.1;  
               08.05.3;  
               08.09.2;  
               08.13.2;  
	       08.22.3)} 

   \title{Spectroscopy of the candidate luminous blue variable at the
center of the ring nebula G79.29+0.46}

\author{R.H.M.~Voors
        \inst{1,2}\thanks{Present address:KNMI (Royal Dutch Meteorological
	Institute), Sectie Astmospherische Samenstelling, Postbus 201, 
	NL-3730 AE De Bilt, The Netherlands}
        \and
	T.R.~Geballe\inst{3}
	\and
        L.B.F.M.~Waters\inst{4,5}
        \and
        F.~Najarro\inst{6}
        \and
        H.J.G.L.M.~Lamers\inst{1,2}
        }   

 \institute{Astronomical Institute, University of Utrecht,
 Princetonplein 5, 3508 TA Utrecht, The Netherlands
        \and
 SRON Laboratory for Space Research, Sorbonnelaan 2,
  NL-3584 CA Utrecht, The Netherlands
  	\and
  Gemini Observatory, 670 N. A'ohoku Place, Hilo, Hawaii 96720, USA
        \and
  Astronomical Institute 'Anton Pannekoek', University of Amsterdam,
  Kruislaan 403, NL-1098 SJ Amsterdam, The Netherlands
        \and
  SRON Laboratory for Space Research, P.O. Box 800, NL-9700 AV Groningen,
  The Netherlands
        \and
  CSIC Instituto de Estructura de la Materia, Dpto. Fisica Molecular,
  C/Serrano 121, E-28006 Madrid, Spain
 }

   \offprints{L.B.F.M. Waters, rensw@astro.uva.nl}

\date{Received date: 23 March 2000; accepted date}
\titlerunning{The central star of G79.29+0.46}

   \maketitle

   \begin{abstract}

We report optical and near-infrared spectra of the central star of the
radio source G79.29+0.46, a candidate luminous blue variable. The spectra
contain numerous narrow (FWHM~$<$~100 km\,s$^{-1}$) emission lines of
which the low-lying hydrogen lines are the strongest, and resemble spectra
of other LBVc's and B[e] supergiants. A few prominent infrared lines are
unidentified. The terminal wind speed is determined from H$\alpha$ to be
110 km\,s$^{-1}$. The strength of H$\alpha$ implies the presence of a very
dense wind.  Extended emission from H$\alpha$ and [N\,{\sc ii}] was
detected but appears to be associated with the Cygnus X region rather than
the radio source. Both diffuse interstellar bands and interstellar
absorption lines are present in the optical spectrum of the central star,
suggesting that there are both diffuse and molecular cloud components to
the extinction and implying a minimum distance of 1~kpc and minimum
luminosity of $\sim$10$^5$\,L$_{\sun}$ for the star. The new spectra and
their analysis indicate a low excitation, dense, and slowly expanding wind
and support the LBVc classification.

\keywords{Circumstellar matter -- Stars: early-type -- Stars: evolution --
stars: individual: G79.29+0.46 -- Stars: mass-loss --
Stars: variables: general} 
  
\end{abstract}

\section{Introduction}

Many gaps exist in the understanding of the post-main sequence evolution
of the most massive ($>$\,40~M$_{\sun}$) stars, due in large part to the
small number of such stars identified as being in this evolutionary phase.
It is known that during part of the post-main sequence phase, identified
as the Luminous Blue Variable (LBV) stage (Conti 1984), these stars lose a
large amount of mass in a short time interval (e.g. Chiosi \& Maeder
1986). The identifying characteristics of an LBV in addition to its blue
colors are (1) a mass loss rate of ($\sim 10^{-5}$~M$_{\sun}$\,yr$^{-1}$),
(2) a low wind velocity of no more than a few hundred km\,s$^{-1}$, (3)
photometric variations of up to 2 magnitudes on time-scales ranging from
months to decades, and (4) the presence of a circumstellar (and sometimes
bipolar) nebula (Nota \& Clampin 1997).

Less than a dozen confirmed LBVs are known in our Galaxy. However, in
recent years a number of LBV candidates (LBVc's) have been found.  
Candidates generally have both of the first two characteristics listed
above and one of the other two (for reviews see Nota \& Lamers 1997). One
recently discovered LBVc is the central star of the radio shell
G79.29+0.46 (Higgs et al. 1994, hereafter HWL; and Waters et al. 1996,
hereafter Paper~I). The principal evidence for its candidacy are its high
luminosity and mass loss rate (estimated as $\sim$1~$\times$~10$^{-6}$
M$_{\sun}$\,yr$^{-1}$ in Paper I) and the presence of a slightly bipolar
ring nebula. Few spectra have been reported and the only spectral features
in them thought to be associated with the star are several recombination
lines of hydrogen and helium, most notably H$\alpha$ (HWL) and Br$\alpha$
(Paper~I). HWL also reported the presence of diffuse interstellar lines
and bands.

In this paper we present and discuss additional optical and near-infrared
spectra of the central star of G79.29+0.46 (hereafter G79*). The
resolution of the optical spectrum is considerably higher than the
spectrum published by HWL and the infrared spectra provide considerably
more wavelength coverage than the spectrum in Paper~I. The new spectra
allow us to estimate the excitation, luminosity and current activity of
the star. We conclude that although G79* may have gone through a more
intense mass loss episode in the recent past, it probably still is in the
luminous blue variable stage of evolution.

\section{Observations and data reduction}
\subsection{Optical}

A log of observations is provided in Table~\ref{obslog}. High resolution
optical spectra of G79* were obtained in 1995 with the Utrecht Echelle
Spectrograph (UES) mounted on the William Herschel Telescope on La Palma.
The TEK~5 detector (1024$\times$1024 pixels) was used. The slit length was
15 arcsec, allowing a portion of the nebula to be observed simultaneously.
Insufficient time was available to obtain the blue part of the spectrum,
which is very faint due to the high extinction. Two overlapping settings
were chosen to cover the red part of the spectrum, with three exposures
obtained at each setting. A spectrum of a ThAr lamp combined with telluric
lines in the spectrum of the G79* allowed the wavelength calibration
to be determined to an accuracy of 1~km\,s$^{-1}$ and provided an estimate
of $\sim44,000$ for the resolving power.

Data reduction used the Midas echelle package. Because of the faintness of
the object ($V~\approx$~16~mag is about the limiting magnitude for the UES)
and because the position of the orders on the CCD was not constant during
the night, considerable care had to be taken (for details see Voors 1999).
The signal-to-noise ratio of the continuum in the final spectrum (not shown)
increases from $\approx$\,5 around 6000~\AA\ to $\approx$\,50 around
8000~\AA.

\begin{table}
\caption[]{Observing Log}
\label{obslog}
\begin{center}
\begin{tabular}{ccc}
Date & Resolving Power & Range ($\mu$m) \\
            \hline
1995 Aug 3  & 44,000       & 0.54 -- 0.76 \\
1995 Aug 3  & 44,000       & 0.54 -- 0.90 \\
1996 Jul 15 & 2515 -- 2775 & 1.07 -- 1.18 \\
1996 Jul 16 & 2540 -- 2800 & 1.62 -- 1.78 \\
1996 Jul 17 & 1600 -- 1870 & 2.05 -- 2.38 \\
1997 May 8  &  800 -- 1050 & 1.01 -- 1.34 \\
1997 May 8  &  725 -- 1000 & 1.86 -- 2.52 \\
	    \hline
\end{tabular}
\end{center} 
\end{table}   

\subsection{Infrared}

Spectra covering portions of the 1.0-2.5~$\mu$m band were obtained at the
3.8m United Kingdom Infrared Telescope (UKIRT) on Mauna Kea in 1996 and
1997. The facility cold grating spectrometer CGS4 (Mountain et al. 1990)
was used for all observations, with its 80$\arcsec$ slit of width of
1$\farcs$2 oriented east-west. In 1996 a 150 l/mm grating was used,
providing a relatively narrow wavelength coverage and high spectral
resolution; in 1997 a 75 l/mm grating was employed to provide wider
spectral coverage at lower resolution.  The resolving powers,
$\lambda$/$\Delta \lambda$, given in Table~\ref{obslog} increase with
wavelength in each spectral interval and are those of the original
(unreduced) spectra. Observations were made in the standard
stare/nod-along-slit mode, with a nod of $\approx 15 \arcsec$; thus any
constant extended emission (from the nebula or background) along the slit
was subtracted away.

The spectra were wavelength-calibrated using observations of Kr and Ar arc
lamps made nearly simultaneously with the measurements of G79*, were
flux-calibrated and corrected for telluric absorption features using
observations of nearby bright dwarf stars with spectral types from mid to
late F, and then were slightly smoothed (reducing the resolving powers by
$20\%$ from those given in Table~\ref{obslog}). Brackett and Paschen
series absorption lines in clear parts of the spectrum were edited out of
the dwarf star spectra prior to ratioing; the uncertainties in this
procedure lead to significant uncertainties in the strengths of the weak
Brackett series lines in G79* in the $H$ band and to slight uncertainty in
the details of the wings of its Br\,$\gamma$ line. The noise level at most
wavelengths can be estimated from the (small) fluctuations in flat
portions of the spectrum.  Near 1.9~$\mu$m, strong telluric water vapor
absorption is responsible for large systematic fluctuations. The rms
wavelength accuracy is typically $\pm$0.0002~$\mu$m.

The $J$ band spectrum obtained in 1996 with the 150 l/mm grating in third
order was contaminated shortward of 1.105~$\mu$m by continuum emission
near 1.6~$\mu$m in second order; this was corrected assuming no strong
emission lines were present near 1.6~$\mu$m and that the slope of the $J$
band continuum longward of 1.105~$\mu$m extrapolated shortward of
1.105~$\mu$m. These assumptions are borne out by the uncontaminated
spectrum obtained in 1997 with the 75 l/mm grating in second order.   

\begin{figure}
        \epsfxsize=8.8cm \epsfysize=8.2cm \epsfbox{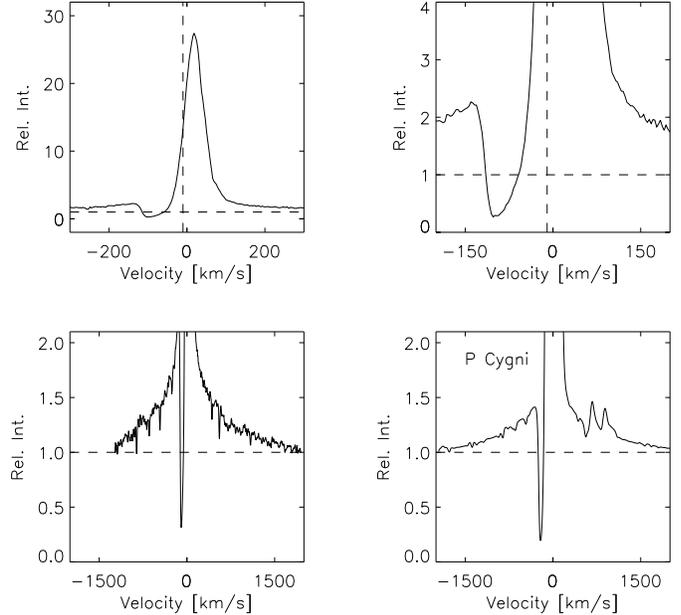}
        \caption[]{The H$\alpha$ line profile
               in G79*, shown in three different velocity and
               intensity scales, and in P Cygni (bottom right)
              }
        \label{fig1}
    \end{figure}

\section{The spectrum of the star and its wind}
\subsection{Optical}

The optical spectrum of G79* contains a number of lines originating in
the stellar photosphere and wind, extended line emission from [N\,{\sc
ii}] and H$\alpha$, and diffuse interstellar bands (DIBs).  
Table~\ref{optlines} lists the identified stellar lines, their
measured central wavelengths, type of profile, and total equivalent width.
Figures 1-4 show many of these lines. Both pure emission and absorption
profiles are observed as well as hybrids.  Some of these lines appear to
be formed entirely in the wind, whereas others show photospheric
contributions.  Because of their wind origins, the hydrogen and helium
lines are unsuitable for determination of the radial velocity of the star.
Instead we use the weak emission lines of metals, whose profiles appear
similar (e.g., Fig.~3) to derive V$_{hel}$~=~-10~$\pm$~2~km~s$^{-1}$
(corresponding to V$_{LSR}$~=~-27~km~s$^{-1}$), a result intermediate
between those of HWL and Paper~I.

\begin{figure}
        \epsfxsize=8.8cm \epsfysize=8.8cm \epsfbox{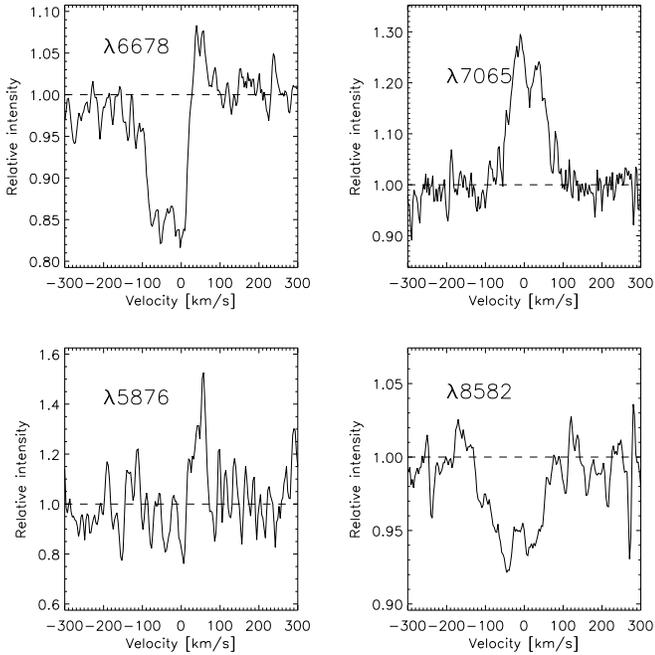}
        \caption[]{Profiles of four He\,{\sc i} lines in G79*.
              } 
        \label{fig2}
    \end{figure}

\begin{figure}
        \epsfxsize=8.8cm \epsfysize=8.5cm \epsfbox{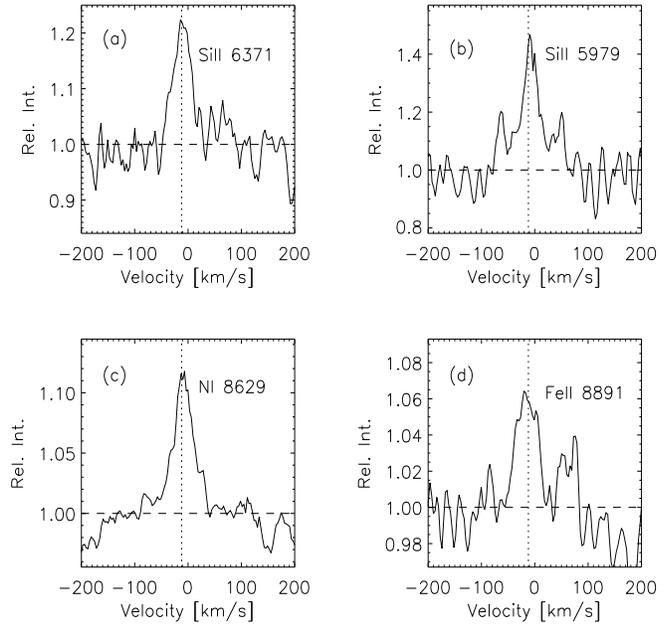}
        \caption[]{Representative metallic lines in G79*.
	The dashed vertical line corresponds to the derived stellar
	radial velocity of -10~km~s$^{-1}$ (hel.).
              } 
        \label{fig3}
    \end{figure}

\begin{table*}
\caption[]{Optical Lines}
\label{optlines}
\vspace{0.5cm}
\small
\begin{flushleft}
\begin{tabular}{lllllc}
$\lambda_{cen}$ [\AA]$^{1}$ & $\lambda_{lab}$ [\AA]$^{1}$ &  [Elem.] & Trans. + multiplet &
Type$^{2}$ & EW [\AA]$^{3}$ \\
\hline
6563.1  & 6562.797 & H\,{\sc i}         & 2-3 (H$\alpha$) (1)   & pc            & 51    \\
8598.1  & 8598.396 & H\,{\sc i}         & 3-15 (Pa\,14) (9)     & ph + pc       & $-0.40$       \\
8545.0  & 8545.387 & H\,{\sc i}         & 3-14 (Pa\,15) (10)    & ph + pc       &       \\
8582.1  & 8582.670 & He\,{\sc i}        & 3p3P - 10d3D          & ph            & $-0.28$       \\
        & 8581.856 & He\,{\sc i}        & 3d3D - 14f3F          & ph            &       \\
6679.1  & 6678.154 & He\,{\sc i}        & 2p1P - 3d1D (46)      & ph + pc       & $-0.41$       \\
7065.2  & 7065.176 & He\,{\sc i}        & 2p3P - 3s3S (10)       & em           & 0.56  \\
6577.9  & 6578.052 & C\,{\sc ii}        & 3s2S - 3p2P (2)         & abs         & ($-0.2$)      \\
6582.6  & 6582.882 & C\,{\sc ii}        & 3s2S - 3p2P (2)         & abs         & ($-0.1$)      \\
7231.0  & 7231.333 & C\,{\sc ii}        & 3p2P - 3d2D (3)         & em          & (0.1) \\
7236.1  & 7236.420 & C\,{\sc ii}        & 3p2P - 3d2D (3)         & em          &       \\
8628.9  & 8629.235 & N\,{\sc i}         & 3s2P - 3p2P (8)         & em          & 0.12  \\
        & 8594.000 & N\,{\sc i}         & 3s2P - 3p2P (8)         & em          &       \\
7423.2  & 7423.641 & N\,{\sc i}         & 3s4P - 3p4S (3)         & em          &       \\
7441.9  & 7442.298 & N\,{\sc i}         & 3s4P - 3p4S (3)         & em          & 0.02  \\
7468.1  & 7468.312 & N\,{\sc i}         & 3s4P - 3p4S (3)         & em          & (0.05)    \\
6481.6  & 6482.048 & N\,{\sc ii}        & 3s1P - 3p1P (8)         & abs 	& $-0.23$    \\
5675.7  & 5676.017 & N\,{\sc ii}        & 3s3P - 3p3D (3)         & abs 	&       \\
5679.4  & 5679.558 & N\,{\sc ii}        & 3s3P - 3p3D (3)         & abs 	& $-0.20$:      \\
5686.0  & 5686.213 & N\,{\sc ii}        & 3s3P - 3p3D (3)         & abs 	& $-0.17$       \\
6402.1  & 6402.246 & Ne\,{\sc i}        & 3s1  - 3p1  (1)         & abs 	& $-0.10$       \\
8234.3  & 8234.636 & Mg\,{\sc ii}       & 4p2P - 5s2S (7)         & em  	&       \\
7056.4  & 7056.712 & Al\,{\sc ii}       & s4s3S - s4p*3P (3)      & em  	& 0.14  \\
7471.1  & 7471.410 & Al\,{\sc ii}       & s3d1D - s4f*1F (21)     & em  	& 0.04  \\
6836.8  & 6837.128 & Al\,{\sc ii}       & s4p*3P - s5s3S (9)      & em  	& 0.02  \\
6231.5  & 6231.750 & Al\,{\sc ii}       & s4p*3P - s4d3D (10)     & em  	& 0.19  \\
6226.0  & 6226.195 & Al\,{\sc ii}       & s4p*3P - s4d3D (10)     & em  	& (0.1) \\
5957.3  & 5957.559 & Si\,{\sc ii}       & 4p2P - 5s2S (4)         & em  	& (0.1) \\
5978.7  & 5978.930 & Si\,{\sc ii}       & 4p2P - 5s2S (4)         & em  	& 0.36: \\
6371.0  & 6371.371 & Si\,{\sc ii}       & 4s2S - 4p2P (2)         & em  	& 0.21  \\
7462.1  & 7462.654 & Si\,{\sc iii}      & s4d3D - s5p*3P          & em  	& 0.07  \\
7466.0  & 7466.321 & Si\,{\sc iii}      & s4d3D - s5p*3P          & em  	& (0.05)  \\
7512.9  & 7513.162 & Fe\,{\sc ii}       & (5D)5s e6D - (5D)5p w6P & em  	& 0.06  \\
7731.3  & 7731.673 & Fe\,{\sc ii}       & (5D)5s e6D - (5D)5p w6P & em  	& 0.05: \\
8890.6  & 8890.899 & Fe\,{\sc ii}    	& (3F2)4p y4F - 4s2 4F    & em  	& 0.11  \\
8897.8  & 8897.810 & Fe\,{\sc ii}	& (3F2)4p y4F - 4s2 4F    & em  	& 0.06   \\
8926.4  & 8926.635 & Fe\,{\sc ii}       & (5D)5s e4D - (5D)5p 4D  & em  	& 0.09: \\
\hline
\end{tabular}
\end{flushleft}
$^{1}$ In air \\
$^{2}$ pc = P Cygni; em = emission line; abs = absorption line (in stellar wind); ph = photospheric \\
$^{3}$ Colons denotes uncertainties of 20-50 percent; values in parentheses are marginal detections, 
uncertain by at least 50 percent; no value indicates contamination by
telluric absorption; all other
uncertainties are less than 20 percent.
\normalsize   
\end{table*}

\begin{figure}
        \epsfxsize=8.8cm \epsfysize=18.0cm \epsfbox{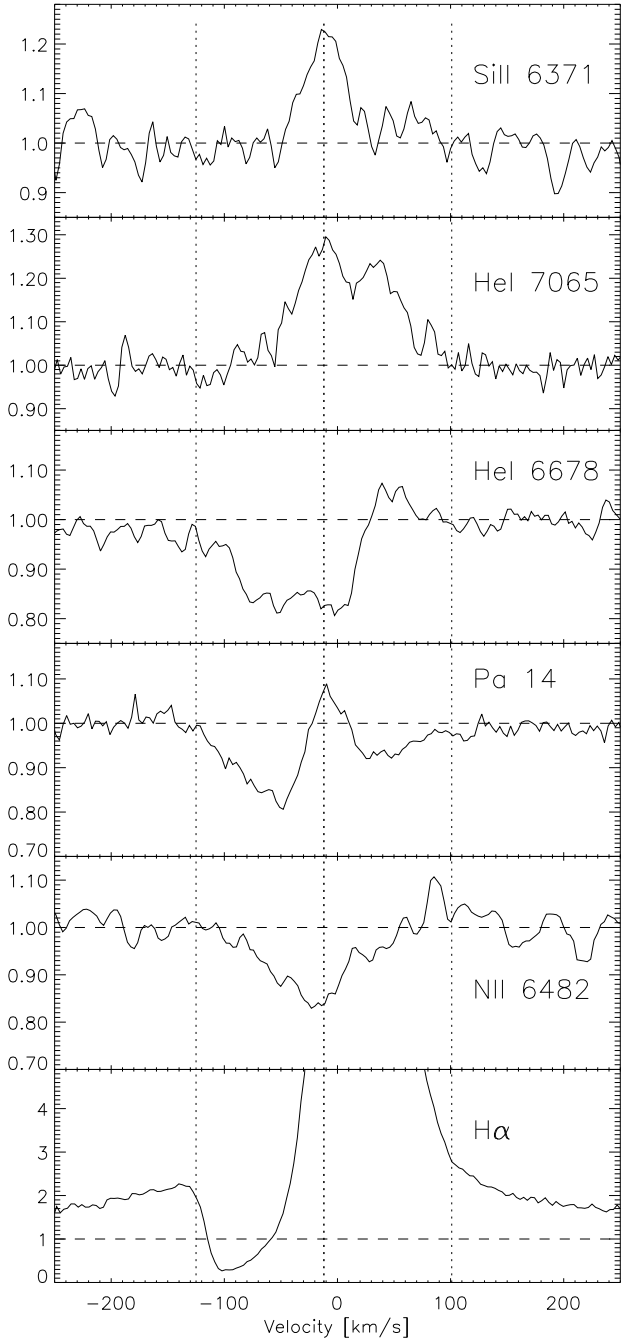}
        \caption[]{Representative line profiles in G79*.
        The dashed vertical lines correspond to the derived
        radial velocity of -10~km~s$^{-1}$ (hel.) and wind speed of
        110~km~s$^{-1}$.
              }
        \label{fig4}
    \end{figure}

\subsubsection{Hydrogen lines}

The H$\alpha$ line (Fig. 1), which dominates the optical spectrum, has a
strong central emission peak, a blueshifted absorption component, and
broad and shallow wings. Contributions by nebular emission to each of
these are negligible. The overall profile is similar to that seen in other
luminous blue supergiants. In P~Cygni the broad and shallow wings to
H$\alpha$ are believed to be caused by the scattering of photons by
electrons (M\"{u}nch 1948, Bernat \& Lambert 1978). We make the same
interpretation for G79*, as suggested in Paper~I, concurring that the wind
velocity of 1400~km~s$^{-1}$ derived by HWL is erroneous. The wings in
G79* and P~Cygni are approximately the same width, but those in G79* are
about twice as strong.  Scattering wings are not evident in the Br$\alpha$
profile in Paper~I. 

Despite the strong similarity of the H$\alpha$ profiles in G79* and
P~Cygni, direct comparison of the profiles does not give a useful estimate
of the mass loss rate from G79*. The strength of scattering wings depends
on the relative strengths of the line and continuum radiation fields and
the electron scattering optical depth in the region where the line forms.
These in turn are determined by the rate of mass loss, the velocity field,
and the effective temperature of the star. The uncertainty in the distance
to G79* (1 to 4kpc, see below) implies a factor of 8 uncertainty in the
mass loss rate. This uncertainty is further increased if the two stars
have different velocity fields. Moreover, even if the same radius and
velocity structure are assumed for both objects a direct comparison would
still fail due to the remarkably different ionizations of P~Cygni and
G79*. The weakness of the He~I lines as well as the presence of N~I and
O~I lines in the G79* spectrum denote a lower effective temperature for
this object than P~Cygni. A difference in the effective temperature is
translated not only into a difference in the radiation field in the line
(which affect the electron scattering wings), but also into a difference
in the run of the departure coefficients which control the overall line
emission. Although the strength of H$\alpha$ in G79* indicates the
presence of a very dense stellar wind, an accurate estimate of the mass
loss rate can be achieved only through detailed quantitative analysis.

Although the H$\alpha$ line profile is unsuitable for determining the
stellar radial velocity, the blueshifted absorption feature allows an
accurate determination of the terminal wind speed (Fig.~4).  The short
wavelength edge of this absorption occurs at V$_{hel}$ =
-120~km\,s$^{-1}$, implying a wind speed of 110~km\,s$^{-1}$ (if
turbulence in the wind is small). This compares with the value of
94~km\,s$^{-1}$ obtained in Paper~I based on the FWZI of the lower
signal-to-noise ratio Br$\alpha$ line profile, which has only a weak
blueshifted absorption feature. On the positive velocity side of the
profile, a change in slope is evident at V$_{hel}$ = +100~km\,s$^{-1}$,
beyond which the scattering wing dominates, in good agreement with the
newly derived wind speed.

The Pa\,14 profile, which also is shown in Fig.~4, is considerably 
different from H$\alpha$.  The emission component, seen near line center 
is formed in the wind, where non-LTE effects tend to overpopulate the high 
levels. The absorption wings are formed in the photosphere.

\subsubsection{Helium lines} 

Four helium lines were detected in the UES spectrum and are shown in
Fig.~2. The He\,{\sc i} 7065 profile shows two peaks, centered at $-12$
km\,s$^{-1}$ and $+37$~km\,s$^{-1}$ (hel.). The profile shown is
corrected for telluric H$_{2}$O lines, and the dip near $v = +10$
km\,s$^{-1}$ is real. The blue peak at $-12$ km\,s$^{-1}$ coincides with
the system velocity, suggesting that it is formed near the star, where
the outflow velocity is low. If the second peak at +37~km\,s$^{-1}$ is
formed in a rotating disk, in matter ejected by the star, or in a
companion star, it is not clear why this feature would not be present in
lines of other elements. In contrast, the He\,{\sc i}\,6678 line has a
P~Cygni profile.  Assuming G79* to be spherically symmetric, the
presence of some absorption redshifted from the stellar velocity of 
-10 km s$^{-1}$ indicates a substantial photospheric contribution.  
However, it is not clear why He\,{\sc i}\,6678, which has a larger
oscillator strength than He\,{\sc i}\,7065 shows the larger photospheric
contribution.

The He\,{\sc i}\,5876 line peak occurs at $+40$ km\,s$^{-1}$, coinciding
with one of the two peaks seen in the $\lambda$\,7065 and $\lambda$\,6678
profiles. However, the stronger component at $-12$ km\,s$^{-1}$ is not
seen. In summary, the helium lines in G79* show large differences from
one another, most likely due to non-LTE effects and optical depth effects
in the photosphere and wind, but we do not have detailed explanations for
their specific disparate behaviors.

\subsubsection{Metal lines}

Among the metallic lines in the optical spectrum of G79* the N\,{\sc
ii} lines, the C\,{\sc ii} lines and the Ne\,{\sc i} 6402 line are in
absorption; all others are in emission. A number of emission line profiles
are shown in Fig.~3; these and others have been used to estimate the
radial velocity of G79*, as discussed previously. Those of metals have
FWHMs of approximately 50 km\,s$^{-1}$ and FWZIs of 90 to 100
km\,s$^{-1}$.  The excitations of the metal lines are low and are typical
for mid to late-B type supergiants.

\begin{figure*}
        \epsfxsize=17.6cm \epsfysize=16.0cm \epsfbox{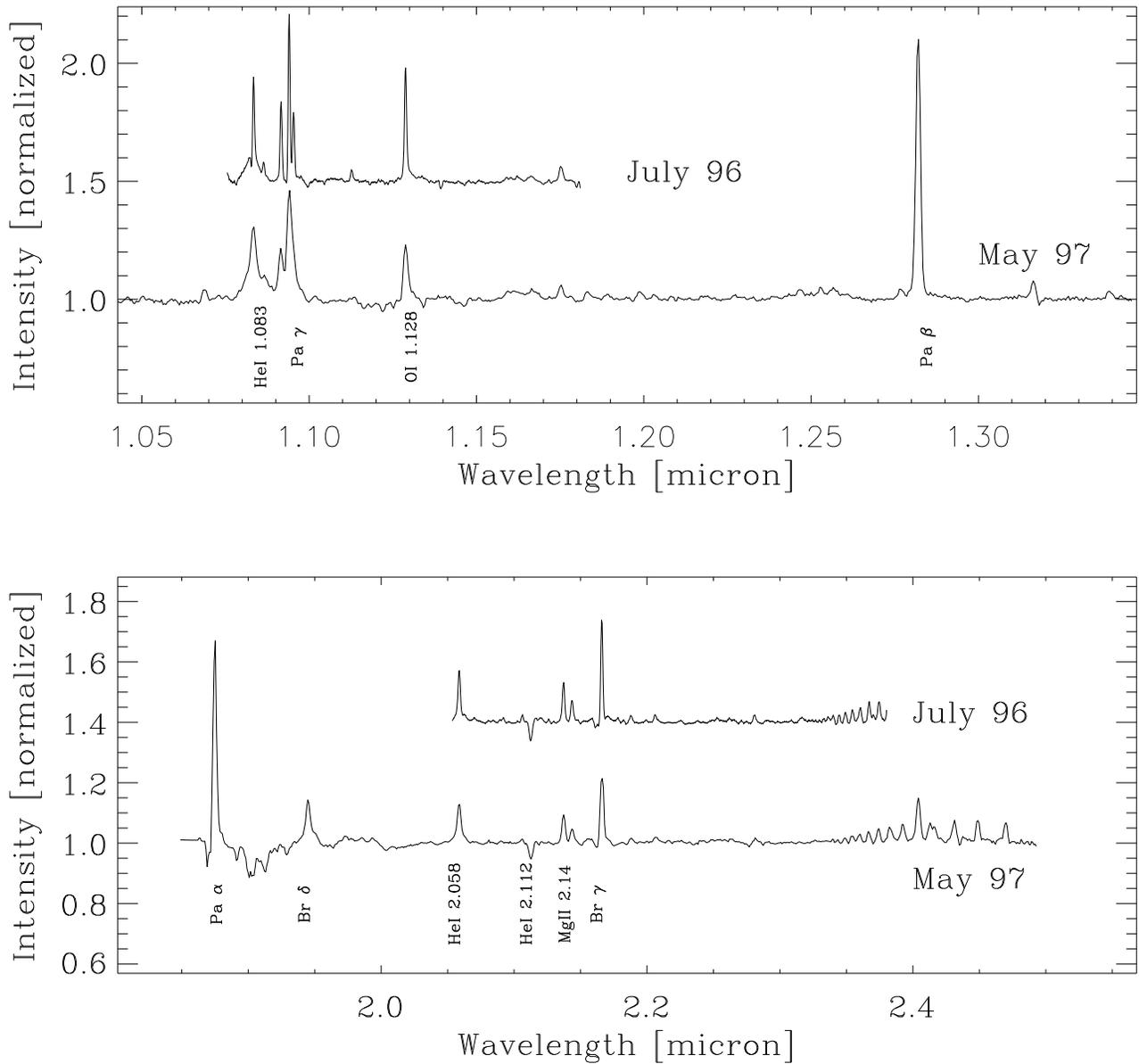}
        \caption[]{Infrared spectra of G79*.
              }
        \label{fig5}
    \end{figure*}

\subsection{Infrared}

The $J$ and the $K$ band spectra of G79* are shown in Fig.~5, with
identifications and equivalent widths given in Tables 3 and 4. Most of the
strong lines are standard recombination lines of hydrogen and helium, but
there are a few unidentified lines of surprising strength. Where a line
was observed more than once, the variation in its equivalent width was
less than 10 percent. Furthermore, after convolving the 1996 spectra to
the resolution of the 1997 spectra, no significant differences in line
shape or width were found.

\begin{table}
\caption[]{Identified Infrared Lines}
\label{ir}
\begin{flushleft}
\begin{tabular}{lllll}
$\lambda_c^{96}$ [\AA]$^{1}$ & $\lambda_c^{97}$ [\AA]$^{1}$ & $\lambda_{lab}$ 
[\AA]$^{1}$ &    Transition & EW [\AA]$^{2}$\\
\hline
        & 10687.5 & 10688.1     & C\,{\sc i}            & 0.8 \\  
10833.7 & 10832.9 & 10833.0     & He\,{\sc i}           & 3.2 \\
10863.8 & 10866   & 10862.6     & Fe\,{\sc ii}          & 0.3 \\
10940.1 & 10940.7 & 10941.1     & Pa\,$\gamma$          & 4.8 \\
11127.3 &         & 11125.6     & Fe\,{\sc ii}          & 0.4 \\
11288.0 & 11288.3 & 11289.4     & O\,{\sc i}            & 3.9 \\
11663   & 11665   & 11664.1     & C\,{\sc i}            & 0.5 \\
11752.5 & 11752.0 & 11752.7     & C\,{\sc i}            & 0.8 \\ 
        & 11830   & 11831       & [Fe\,{\sc ii}]        & 0.7 \\
        & 12466.1 & 12466.2     & N\,{\sc i}            & 0.5 \\
        & 12566.5 & 12567.0     & [Fe\,{\sc ii}]        & 0.5 \\   
        & 12820.1 & 12821.6     & Pa\,$\beta$           & 19.3 \\
        & 13164.1 & 13164.6     & O\,{\sc i}            & 1.3 \\
\vspace{-0.2cm} \\
16400   &         & 16411.7     & Br\,12                & - \\
16438.9 &         & 16440.0     & [Fe\,{\sc ii}]        & - \\
16875   &         & 1.68        & Fe\,{\sc ii}          &  0.3 \\
16800   &         & 16811.1     & Br\,11                & - \\
16978.5 &         & 16976.1     & Fe\,{\sc ii}          & $<$0.2 \\
17007.1 &         & 17007.0     & He\,{\sc i}           & 0.4  \\
17109.3 &         & 17115.9     & [Fe\,{\sc ii}]        & 0.5 \\
17370   &         & 17366.6     & Br\,10                & - \\
17455.7 &         & 17454.1     & [Fe\,{\sc ii}]        & 1.3 \\
\vspace{-0.2cm} \\
        & 18748.9 & 18756.1     & Pa$\alpha$            & - \\
        & 19450.5 & 19450.9     & Br$\delta$            & 5.9 \\
20587.0 & 20586.5 & 20586.9     & He\,{\sc i}           & 4.4 \\
21125.2 & 21127.6 & 21125.8     & He\,{\sc i}           & $-$1.6; blend \\
        &         & 21139.0     & He\,{\sc i}           & blend \\
21372.6 & 21373.3 & 21374.8     & Mg\,{\sc ii}          & 3.0 \\
21436.4 & 21436.8 & 21438.0     & Mg\,{\sc ii}          & 1.7 \\
21609.9 &         & 21613.7     & He\,{\sc i}           & -0.7 \\
21659.7 & 21662.3 & 21661.2     & Br\,$\gamma$          & 7.4 \\
22061.4 & 22067.2 & 22062.4     & Na\,{\sc i}           & 0.7 \\
22081.9 &         & 22089.7     & Na\,{\sc i}           & 0.2 \\   
23191.1 &         & 23195.6     & Pf\,38                & $<$0.2 \\
23217.1 &         & 23218.0     & Pf\,37                & $<$0.2 \\
23242.1 &         & 23242.4     & Pf\,36                & $<$0.2 \\
23268.0 &         & 23268.9     & Pf\,35                & $<$0.2 \\   
23295.1 &         & 23297.9     & Pf\,34                & $<$0.2 \\
23327   &         & 23329.6     & Pf\,33                & 0.2 \\
23361.8 &         & 23364.4     & Pf\,32                & 0.3 \\
23400.3 & 23396.1 & 23402.8     & Pf\,31                & 0.5 \\
23445.3 & 23433.8 & 23445.3     & Pf\,30                & 0.5 \\
23492.3 & 23487.5 & 23492.4     & Pf\,29                & 0.7 \\
23545.6 & 23543.4 & 23544.8     & Pf\,28                & 0.9 \\
23604.5 & 23599.2 & 23603.5     & Pf\,27                & 1.2 \\
23670.6 & 23667.4 & 23669.4     & Pf\,26                & 1.4 \\
23746.7 & 23742.0 & 23743.8     & Pf\,25                & 1.6 \\
        & 23829.8 & 23828.2     & Pf\,24                & 2.2 \\
        & 23925.0 & 23924.7     & Pf\,23                & 2.5\\ 
        &         & 24035.5     & Pf\,22                & blend \\
        & 24041.2 & 24041.5     & Mg\,{\sc ii}          & blend \\
        &         & 24044.5     & Mg\,{\sc ii}          & 5.8; blend \\
        & 24131.7 & 24124.6     & Mg\,{\sc ii}          & 3.0 \\
        & 24159.7 & 24163.9     & Pf\,21                & 2.7 \\
        & 24312.0 & 24313.6     & Pf\,20                & 2.9 \\ 
        & 24488.1 & 24490.0     & Pf\,19                & 2.8 \\
        & 24696.7 & 24952.5     & Pf\,18                & $\approx$2.5\\
\hline
\end{tabular}
$^{1}$ In vacuo\\
$^{2}$ EW's of H band hydrogen lines highly uncertain and not given.
\end{flushleft}
\end{table}

\subsubsection{Identified lines}

Several of the prominent infrared lines have profiles worthy of note. The
He\,{\sc i} 1.083 and 2.058~$\mu$m lines possess broad wings, which extend
approximately 1,000 km\,s$^{-1}$ to either side of line center. The wings
are somewhat broader in the 1.083~$\mu$m line profile. As in the case of
H$\alpha$, these wings are the result of electron scattering. The weak
absorption feature just shortward of Br\,$\gamma$ (2.166~$\mu$m) is due to
absorption by He\,{\sc i} 7-4, whose strongest transitions are at
2.165~$\mu$m and 2.161~$\mu$m (e.g., see Najarro et al. 1994). The
equivalent width of this absorption is consistent with the much stronger
He\,{\sc i} 4-3 absorption at 2.112~$\mu$m. The O\,{\sc i} line at
1.128~$\mu$m displays a long wavelength shoulder, but this may be due to
blending with lines of Si\,{\sc iii}.

Prominent in the $K$ band are two Mg\,{\sc ii} lines near 2.14~$\mu$m.  
This pair is excited by UV fluorescence. Whether this is continuum or
Ly\,$\beta$ fluorescence (Bowen 1947) cannot be determined from the
present data. The Na\,{\sc i} emission lines at 2.206~$\mu$m and
2.209~$\mu$m also may be excited by fluorescence (Thompson \& Boroson
1977) via 3303 \AA\ (continuum) photons. Both the Mg and Na doublets are
present in the spectra of LBV and B[e] stars (McGregor et al.  1988). In
G79* and Wra 751 the Mg\,{\sc ii} lines are much stronger than the
Na\,{\sc i} lines; however, in AG Car the doublets are similar in strength
(Morris et al. 1996).

The $J$ band lines of O\,{\sc i} at 1.129~$\mu$m and 1.317~$\mu$m form a
third diagnostic infrared pair. The 1.129~$\mu$m line is pumped by
Ly\,$\beta$, and is only strong relative to the 1.317~$\mu$m line if Lyman
fluorescence is the dominant source of excitation (Grandi et al. 1975). In
G79* this ratio is 8.8~$\pm$~0.8, implying a strong Ly\,$\beta$ flux,
as expected.

Pfund series lines are detected in G79* from n=18 to n=37.
Observations of higher n lines are made difficult by blending and by the
presence of an unidentified line at 2.317~$\mu$m. High Pfund series lines
also are found in Be stars (e.g. $\gamma$~Cas; Hamann \& Simon 1987).

\begin{table}
\caption[]{Unidentified Infrared Lines}
\label{iruid}
\vspace{0.5cm}
\begin{flushleft}
\begin{tabular}{llll}
$\lambda_c^{96}$ [\AA]$^{1}$ & $\lambda_c^{97}$ [\AA]$^{1}$ & EW [\AA] & Remarks \\
\hline
10916.0 & 10915.8       & 2.5   & He\,{\sc i} + Mg\,{\sc ii} ?\\
10953.3 &               & 2.3   & \\
        & 12768.7       & 1.1   & He\,{\sc i} ?\\
        & 13391         & 0.7   & \\
\vspace{-0.2cm} \\
21062.0 & 21061.3       & 0.5   & Fe\,{\sc ii} ??\\
21879.3 & 21882.0       & 0.6   & Fe\,{\sc ii} ??\\
22808.4 &               & 0.5   & N\,{\sc ii} ??\\
23163.2 &               & 0.5   & also in AG~Car and Wra~751\\
23703.7 &               & 0.6   & \\
\hline
\end{tabular}
$^{1}$ In vacuo
\end{flushleft}
\end{table}

\subsubsection {Unidentified Lines}

A number of unidentified infrared lines are observed in G79* and
listed in Table~\ref{iruid}. The most prominent of these are at 1.0916 and
1.0953 $\mu$m. To our knowledge, neither has been seen previously in any
other objects. A possible identification for the 1.0916~$\mu$m line is the
He\,{\sc i} 3D-3Fo transition. Also possible are two Mg\,{\sc ii} 2D-2Po
transitions, but they have much lower log\,$gf$ values than lines of the
same element not seen in the spectrum. A forbidden [Fe\,{\sc ii}]
transition at 1.0957~$\mu$m is shifted slightly too far from the
1.0953~$\mu$m line to be a serious possibility; moreover a second
[Fe\,{\sc ii}] line at 1.1068~$\mu$m would be present. The line at
1.2769~$\mu$m, just on the blue side of Pa\,$\beta$, is possibly due to
[Fe\,{\sc ii}]. However, the 1.3098~$\mu$m line of the same multiplet
should also be present, but is not observed.

\subsection{Comparison of G79* with massive supergiants}

The optical--infrared spectrum of G79* most closely resembles those of
mid to late type B supergiants. The excitation of the metal lines is
typical of such stars. Note, however, that few B-type supergiants have
N\,{\sc i} emission lines in their optical spectra. Those that do, e.g.
HD~326823 (early B), CPD$\,-52\degr$9243 (B3), HD~316285 (P~Cygni type)
(Lopes et al. 1992), are classified as B[e] supergiants. The existences of
disks in these sources are inferred from double-peaked line profiles (e.g.
HD~326823) or from thermal dust emission (e.g. CPD$\,-52\degr$9243).
Clearly G79* cannot be such a star, as not only are no forbidden lines
detected in its optical spectrum, but also virtually all of its lines lack
double-peaked profiles. The narrow hydrogen emission lines seen by us and
in Paper~I imply massive low velocity winds, similar to those seen in
LBVs. This, plus the presence of a circumstellar, dusty nebula suggest
that the star is indeed an LBV, even if it is currently not variable.

\begin{figure}
        \epsfxsize=8.8cm \epsfysize=6.6cm \epsfbox{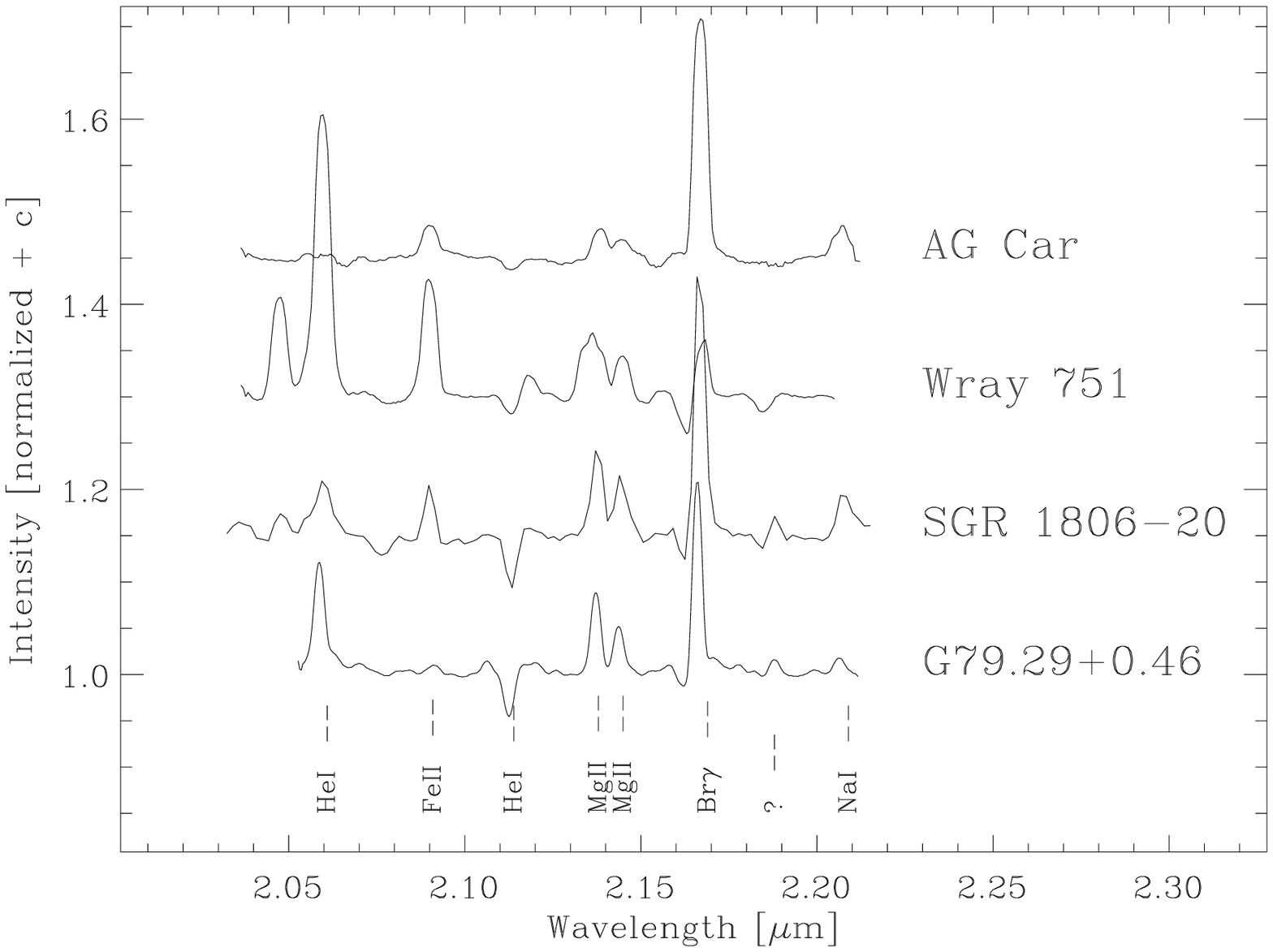}
        \caption[]{Comparison of K-band spectra of G79*, the LBVs 
	Wra~751 and AG~Car, and the SGR~1806-20, which is dominated by
	the B supergiant companion. 
              }
        \label{fig6}
    \end{figure}

In Fig. 6 the $K$ band spectra of G79* and three massive supergiants,
SGR~1806$-$20 (van Kerkwijk et al. 1995), Wra~751 and AG~Car (Morris et
al. 1996), are compared. Wra~751 and AG~Car are well studied LBVs. The $K$
band spectrum of SGR~1806$-$20 is thought to be dominated by a B
supergiant whose optical position is close to that of the soft gamma
repeater source; note that there is some doubt as to the physical
association between the gamma ray source and the B supergiant (Fuchs et
al. 1999). The similarity between the spectra of G79* and this star is
striking, the main difference being the strength of the (permitted)
Fe\,{\sc ii}\,2.089~$\mu$m line. As noted by Morris et al. (1996), the
near-infrared spectra of many types of massive evolved stars, are similar;
thus one cannot conclude based on their similarity that G79* and the
B supergiant in SGR~1806-20 are similar objects.

\section{DIBs, interstellar reddening, and luminosity}
\label{ext} 

Table~\ref{dibs} lists the diffuse interstellar absorption bands present in
the spectrum of G79*. By comparing the colors of G79* with those of
Kurucz (1979) models, HWL derived $A_V$~=~16~mag. We use a different
technique, adopting the extinction law for the Cygnus OB2 association
(Torres-Dodgen et al. 1991), since that region is probably most
representative of the interstellar medium towards G79.29+0.46.  Dereddening
the $JHK$ spectra as well as the optical spectra published by HWL, and
assuming a blackbody temperature for G79* of 18,000~K, results in a
derived $E(B-V)$ of 4.9 $\pm$ 0.4. With an $R_V$ of 3.04 this gives an $A_V$
of 14.9 $\pm$ 1.2, which is slightly less than HWL. A lower blackbody
temperature for the star, which may be more realistic, would result in a
lower derived extinction. In a recent study of the infrared properties of
G79.29+0.46 and its star Trams et al. (1999) have derived an $E(B-V)$ of
3.9, corresponding to $A_V$~=~11.9~mag.

\begin{table}
\caption{Diffuse Interstellar Bands}
\label{dibs}   
\vspace{0.5cm}
\begin{flushleft}
\begin{tabular}{lll}
$\lambda_{c}$ [\AA]     & EW [m\AA]     & Remarks\\
\hline
8621                    & 585 $\pm$ 50  & \\
7927                    & -             & broad feature \\
7721                    & 60 $\pm$ 20   & \\
7585                    & $<$ 30        & \\
7581                    & $<$ 30        & \\
7561                    & 85 $\pm$ 35   & \\
7559                    & 40 $\pm$ 20   & \\
7432                    & -             & broad feature \\
7334                    & -             & tell. contamination \\
6993                    & -             & tell. contamination \\
6940                    & -             & broad feature \\
6661                    & 41 $\pm$ 10   & \\
6379                    & 150 $\pm$ 15  & \\
6284                    & 1500 $\pm$ 150& \\
6281                    & 400 $\pm$ 100 & \\
6270                    & 150 $\pm$ 30  & \\
6204                    & 300 $\pm$ 100 & \\
6196                    & 200 $\pm$ 70  & \\
5797                    & 260 $\pm$ 40  & \\
\hline
\end{tabular}
\end{flushleft}
\end{table}

The strength of the DIBs are a good measure of the interstellar extinction
due to diffuse matter. In several OB associations in Cygnus a tight
relation exists between the strength of the $\lambda$5797 DIB and
$E(B-V)$, (Chlewicki et al., 1986). Assuming this relation for
G79.29+0.46, which is very close in the sky to Cyg~OB2, we obtain $E(B-V)$
= 2.1 $\pm$ 0.3. This is much less than the values derived above. The
discrepancy may be resolved if additional reddening is present in the form
of molecular clouds. G79.29+0.46 is only 9' distant from the center of the
obscured star-forming region DR 15, which is thought to contain a few hot
stars (Odenwald et al., 1990). An IR spectrum obtained with ISO-SWS Morris
et al. (in preparation) shows bands of H$_{2}$O and CO$_{2}$ ice. It
remains to be determined whether the extinction derived from the ISO
spectrum accounts for the difference between the two values of $E(B-V)$,
but it seems safe to conclude that the extinction to G79.29+0.46
contains both diffuse and molecular contributions.

From the radial velocity of G79* and after allowing for peculiar
motions of up to 10~km~s$^{-1}$, we derive a kinematic distance of 2~kpc
from the galactic rotation curve with an uncertainty of approximately a
factor of two (see, e.g., HWL Fig.~14). Trams et al. use their best
fitting model atmosphere, with a luminosity of
3$\times$10$^{5}$~L$_{\sun}$ and mass loss rate of
1.3$\times$10$^{-5}~$M$_{\sun}$ (considerably higher than the estimate of
Paper~I) to derive a distance of 1.8~kpc. It is virtually certain that the
distance exceeds 0.7~kpc since no interstellar clouds closer than that
distance are known which could cause the large observed extinction. The
DR~15 complex, which we have suggested is responsible for some of the
extinction to G79.29+0.46, is 1~kpc distant. We conclude that the distance
to G79.29+0.46 is in the range 1-4~kpc, with a most likely value of
$\sim$2~kpc.

\begin{figure}
        \epsfxsize=8.8cm \epsfysize=8.0cm \epsfbox{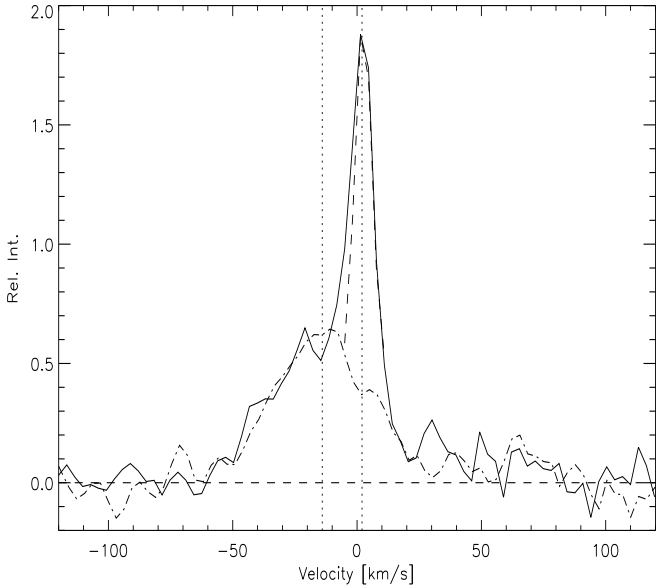}
        \caption[]{Average nebular profiles of H$\alpha$ (solid line) and   
	{\rm [N\,{\sc ii}]}$\lambda$6583 (dash-dotted line), the latter 
	multiplied by a factor of 2.3. The vertical dotted lines denote
	the centers of the {\rm [N\,{\sc ii}]} line and narrow H$\alpha$
        peak. The dashed line overplotted on H$\alpha$ peak indicates the 
	spectral resolution.
              }
        \label{fig7}
    \end{figure}

\section {Extended line emission}

Line profiles, averaged along the 15$\arcsec$ slit, excluding the regions
influenced by stellar radiation, are shown in Fig.~7. The profile of
H$\alpha$ was averaged over a slightly smaller region along the slit than
[N\,{\sc ii}]\,$\lambda$6583, because of contamination by the bright
central peak of stellar H$\alpha$. The widths and the shapes of the broad
components of the [N\,{\sc ii}] and the H$\alpha$ nebular profiles agree
very well, H$\alpha$ being 2.3 times stronger. However, the narrow
nebular component seen in H$\alpha$ is missing in [N\,{\sc ii}]. Clearly
the narrow and broad components originate from different regions.

The broad profiles of both lines are well fitted by a gaussian with FWHM
45$\pm$3~km\,s$^{-1}$. The FWZI is about 70~km\,s$^{-1}$. The central
wavelength of the gaussian is $-14 \pm$ 2~km\,s$^{-1}$ and agrees very
well with the radial velocity of the central star, $-10$~km\,s$^{-1}$.  
More recent observations (Voors 1999) show that both the broad and
narrow components extend well beyond the ring nebula, implying that they
are interstellar in origin and probably associated with emission from the
Cygnus-X region.

\section{Conclusion: the nature of G79*}

Even at the minimum distance of 1~kpc the luminosity of G79* is
1$\times$10$^{5}$~L$_{\sun}$; at the maximum distance of 4~kpc it
approaches that of some of the most luminous stars known. Its temperature
and spectral type are not well defined, but the presence of neutral
species in emission and absorption as well as N II lines in absorption
suggests it is similar to a mid B supergiant. The high luminosity, high
mass loss rate, moderate wind velocity of 110~km~s$^{-1}$, and presence of
a circumstellar nebula all are hallmarks of an LBV. The massive dusty ring
nebula surrounding G79* indicates a period of much higher mass loss in the
recent past (Paper~I), possibly during a red supergiant phase. The star
may evolve into a Wolf-Rayet star. Variability has not been demonstrated,
but given the paucity of observations this is not surprising. More
detailed spectra, such as those recently obtained for the Pistol Star
(Najarro et al. 1998) and FMM362 (Geballe et al. 2000), are needed for
G79* in order to allow quantitative analysis, including a direct estimate
of the mass loss rate, and more careful comparison with other LBVc's and
LBV's.

\begin{acknowledgements}

We thank P. W. Morris for permission to reproduce the $K$ band spectra of
AG~Car and Wra~751, M. van Kerkwijk for the spectrum of SGR~1806-20, and
B. Wolf for helpful comments. We also thank the $\dot{M}$ group in Utrecht
for useful discussions and the referee for helpful comments. LBFMW
acknowledges financial support from a NWO ``Pionier'' grant.  F. N.
acknowledges DGYCIT grants PB96-0883 and ESP98-1351 UKIRT is operated by
the Joint Astronomy Centre on behalf of the U.K. Particle Physics and
Astronomy Research Council. We have made use of the SIMBAD database,
operated at the Centre de Donn\'{e}es Astronomiques, Strasbourg, France.  
\end{acknowledgements}

\end{document}